\documentclass[twocolumn,prb,aps,showpacs,superscriptaddress,floatfix]{revtex4}

\usepackage[dvips]{color,graphicx}

\usepackage{dcolumn}
\usepackage{bm}
\usepackage{latexsym}
\usepackage{amssymb}

\usepackage[colorlinks,bookmarks=false,citecolor=blue,linkcolor=red,urlcolor=blue]{hyperref}

\begin{document}

\title{Time evolution of correlations in strongly interacting fermions after a quantum quench} 

\author{Salvatore R. Manmana}
\affiliation{Institute of Theoretical Physics, \'Ecole
  Polytechnique F\'ed\'erale de Lausanne, CH-1015 Lausanne,
  Switzerland.} 
\author{Stefan Wessel}
\affiliation{Institut f\"ur Theoretische Physik III, Universit\"at
  Stuttgart, Pfaffenwaldring 57, D-70550 Stuttgart, Germany.}  
\author{Reinhard M. Noack}
\affiliation{Fachbereich Physik, Philipps-Universit\"at Marburg, D-35032 Marburg, Germany.}
\author{Alejandro Muramatsu} 
\affiliation{Institut f\"ur Theoretische Physik III, Universit\"at
  Stuttgart, Pfaffenwaldring 57, D-70550 Stuttgart, Germany.}  

\date{\today}


\begin{abstract}
Using the adaptive time-dependent density matrix renormalization group, we 
study the time evolution of density correlations of interacting spinless 
fermions on a one-dimensional lattice after a sudden change in the interaction
strength.
Over a broad range of model parameters, the correlation function exhibits a 
characteristic light-cone-like time evolution 
representative of a 
ballistic transport of information.
Such behavior is observed both when quenching an insulator into the metallic 
region and also when quenching within the insulating region. 
However, when a metallic state beyond the quantum critical point is
quenched deep into the insulating regime, no indication for ballistic transport is observed. 
Instead, stable domain walls in the density correlations emerge during the
time evolution, consistent with the predictions of the  Kibble-Zurek mechanism. 
\end{abstract}

\maketitle

\section{Introduction}
\label{sec:intro}
New experimental possibilities to manipulate ultra-cold atoms with a
high degree of control 
\cite{bloch02,kinoshita06,sadler06,weiler08} have led to a renewed interest in
the dynamics of quantum many-body systems out of equilibrium.
In these systems, the coupling to the environment is negligible, so
that perturbation by external couplings, which would cause relaxation,
does not influence their time evolution.
A particularly simple realization of a time-dependent perturbation is
a so-called quantum quench, in which a 
nonequilibrium situation is induced by suddenly changing one or more
parameters of the system, such as the interaction strength between the particles.
In bosonic systems, experiments have been realized in which a collapse
and revival of the initial state is observed,\cite{bloch02} as well
as others in which relaxation to non-thermal 
states in one dimension,\cite{kinoshita06} where the system becomes virtually 
integrable, takes place.
Of particular interest are quenches through a quantum critical point.
\cite{sadler06,weiler08}
When such quenches are fast enough and end in a phase characterized by spontaneously 
broken symmetry, Kibble \cite{kibble76} and Zurek
\cite{zurek85,zurek96} have predicted the  
spontaneous formation of topological defects.
In this scenario, domains with different realizations of the possible vacua 
of the broken symmetry state
are created, giving rise to topological defects.

Recent theoretical work has investigated the nature of 
quasi-stationary states in correlated quantum 
systems reached a sufficiently long time after
quenches.\cite{rigol07,cazalilla06,kollath07,barthel07,moeckel08,manmana07} 
In addition to studying the long-time behavior, it is also of interest
to provide a physical picture for the 
mechanism of the evolution 
by investigating the short-time behavior of relevant correlation functions. 
In the case where the quench is performed at a critical point, Calabrese and Cardy 
\cite{calabrese06,calabrese07} have provided a general picture based on
renormalization group arguments and, in 
one dimension, on boundary conformal field theory. 
In particular, for a one-dimensional system which possesses
quasiparticle excitations with a typical velocity $v$, they predict
the formation of a light cone, i.e., correlations between points at
distance $x$ are established after time $t \sim x/2v$.
This general picture has been numerically confirmed recently in the Bose-Hubbard model 
using the adaptive time-dependent density matrix renormalization group
(t-DMRG) \cite{vidal03,white04,daley04} and 
exact diagonalization techniques.\cite{laeuchli08}
This system is known to undergo a quantum phase transition from a
superfluid to a Mott-insulating phase.  
The numerical results indicate that a horizon is created irrespective
of the criticality of the system, confirming the picture of Calabrese and Cardy that the
ballistic transport of quasiparticle excitations is the leading
contribution to information transfer in a generic quench.

Here we study a system of spinless fermions on a one-dimensional lattice at 
half filling with a nearest-neighbor repulsion $V$ using the t-DMRG method.
This model is equivalent to the anisotropic Heisenberg (XXZ) chain
and can be obtained from it by applying a Jordan-Wigner transformation.\cite{jordanwigner}
At $V_c=2t_{\text{h}}$, where $t_{\text{h}}$ is the nearest-neighbor
hopping amplitude, the ground-state phase diagram of the model 
contains a quantum critical point separating a Luttinger liquid phase ($V<V_c$)
from a charge-density-wave (CDW) insulator. 
The CDW phase corresponds to a spontaneous breaking of lattice
translational symmetry through the doubling of the unit cell, so
that there are two degenerate sectors for the ground state.  
Hence, for a sufficiently rapid quench into the CDW phase, 
domain walls might be expected to be created, as predicted by the Kibble-Zurek
mechanism.
We carry out quenches starting with initial states in
either phase and suddenly change $V$ to a value that corresponds to
another point in the same phase, to the
other phase, or to the critical point.  
This will allow us to study the extent to which the picture of Calabrese and Cardy is valid,
as well as the details of the evolution of the density correlation
function in 
a regime that goes beyond the scaling theory of the Kibble-Zurek mechanism. 
In addition, we consider the time evolution of the von Neumann entropy
of subsystems of varying size, which 
provides a measure of the spread of entanglement in
the system. \cite{calabrese05,calabrese07a,deChiara05,barmettler08}

The paper is organized as follows.
In Sec.\ \ref{sec:model}, we describe the model, how the quench is
carried out, and some details of the calculations.
In Sec.\ \ref{sec:results}, we present our results for the density 
correlation functions and the block entropy for the various quench scenarios. 
In Sec. \ref{sec:discussion} we summarize and discuss our findings. 

\section{Model and Method}
\label{sec:model}

In the following, we consider interacting spinless fermions on a
one-dimensional lattice at half filling, described by the Hamiltonian
\begin{equation}
\hat{H} = - t_{\text{h}} \sum_j \left( c_{j+1}^{\dagger}
  c_j^{\phantom\dagger} + h.c. \right) 
+ V \sum_j n_j^{\phantom\dagger} n_{j+1}^{\phantom\dagger} \; , 
\label{eq:hamiltonian}
\end{equation}
with  nearest-neighbor hopping amplitude $t_{\text{h}}$ and
nearest-neighbor repulsion $V$. 
The operator $c_i^{\left( \dagger \right)}$ annihilates (creates) a fermion at
lattice site $i$, and  $n_i = c_i^{\dagger} c_i^{\phantom\dagger}$
is the local density operator at site $i$. 
In the following, we take $\hbar = 1$, set the lattice constant $a=1$,  and 
measure energies in units of $t_{\text{h}}$ and  times in units of $1/t_{\text{h}}$. 
We denote the above Hamiltonian for a given value of $V$ as $H(V)$.
The ground-state phase diagram of the integrable quantum system
described by $H(V)$ can be obtained via the Bethe
ansatz\cite{gap_Bethe} and is well known to  
contain a quantum phase transition at $V_c = 2$ from a Luttinger-liquid regime 
for $V < V_c$ to a charge-density-wave (CDW) insulating regime for $V > V_c$,
in which translational symmetry is spontaneously broken in the
thermodynamic limit.
We are interested here in the time evolution after a quantum quench in 
which the initial state of the system is prepared in the
ground state of the Hamiltonian $H(V_0)$ with an initial value of
the interaction set to $V_0 = 0.5$ or $V_0 = 10$, i.e., we consider
quenches starting from both a metallic state and an insulating one.
After the quench, i.e., the sudden change of the interaction
parameter, the dynamics of the
closed quantum system is governed by the Schr\"odinger equation with
Hamiltonian $H(V)$, with a value $V\neq V_0$.  
Quenches that cross the quantum critical point will be treated, as well as 
those that remain within the same phase. 
We calculate the time evolution of the equal-time density correlation function
\begin{equation}
C_{i,j}(t)= \langle n_i(t) n_j(t) \rangle - \langle n_i(t) \rangle \langle
n_j(t) \rangle \, ,
\label{eq:densdens}
\end{equation}
as well as the local density $\langle n_i (t) \rangle$ of the system.
The time evolution of these observables is computed using the adaptive
$t$-DMRG for systems with $L=49$ and, in some cases, $L=50$ lattice
sites, both with $N=25$ particles. 
In choosing an odd number of lattice sites, we ensure a unique 
ground state in the CDW regime, accessible to the DMRG method. 
The results presented for $C_{i,j}(t)$ are computed from the center of
the system.
We confirm explicitly that reflection symmetry 
remains conserved over the time periods considered.  
We  perform the calculations by fixing the discarded
weight to $10^{-9}$, keeping up to several hundred states during the 
time evolution.
In the following, we concentrate on the behavior of the system for times
$t \leq 5$, whereas in a previous publication,\cite{manmana07} we focused 
on  the long-time relaxation behavior after a quantum quench in this 
system.

As an additional probe for the propagation of information through the
system after the quench,  we calculate within the $t$-DMRG the time
evolution of the von Neumann entropy for a subsystem $A$ of $l$ sites, 
\begin{equation}
S_l(t) = - {\text{Tr}}\left[ \varrho_A \log(\varrho_A) \right(t)] \, ,
\label{eq:blockentropy}
\end{equation}
where $\varrho_A(t)$ is the reduced density matrix of the subsystem.
\cite{calabrese04}
The von Neumann entropy in Eq.~(\ref{eq:blockentropy}) is a measure of the 
entanglement of subsystem $A$ with the remainder of the system. 
Its time evolution therefore enables us to quantify the growth of the
entanglement in the quantum system after the quench.\cite{calabrese05,deChiara05,calabrese07a,barmettler08}
In the following sections, we will discuss the propagation of information and 
entanglement after the quantum quench 
based on the time evolution of the quantities described above.

\section{Results}
\label{sec:results}

\subsection{CDW initial state}
\begin{figure*}[t]
\includegraphics[width=\textwidth]{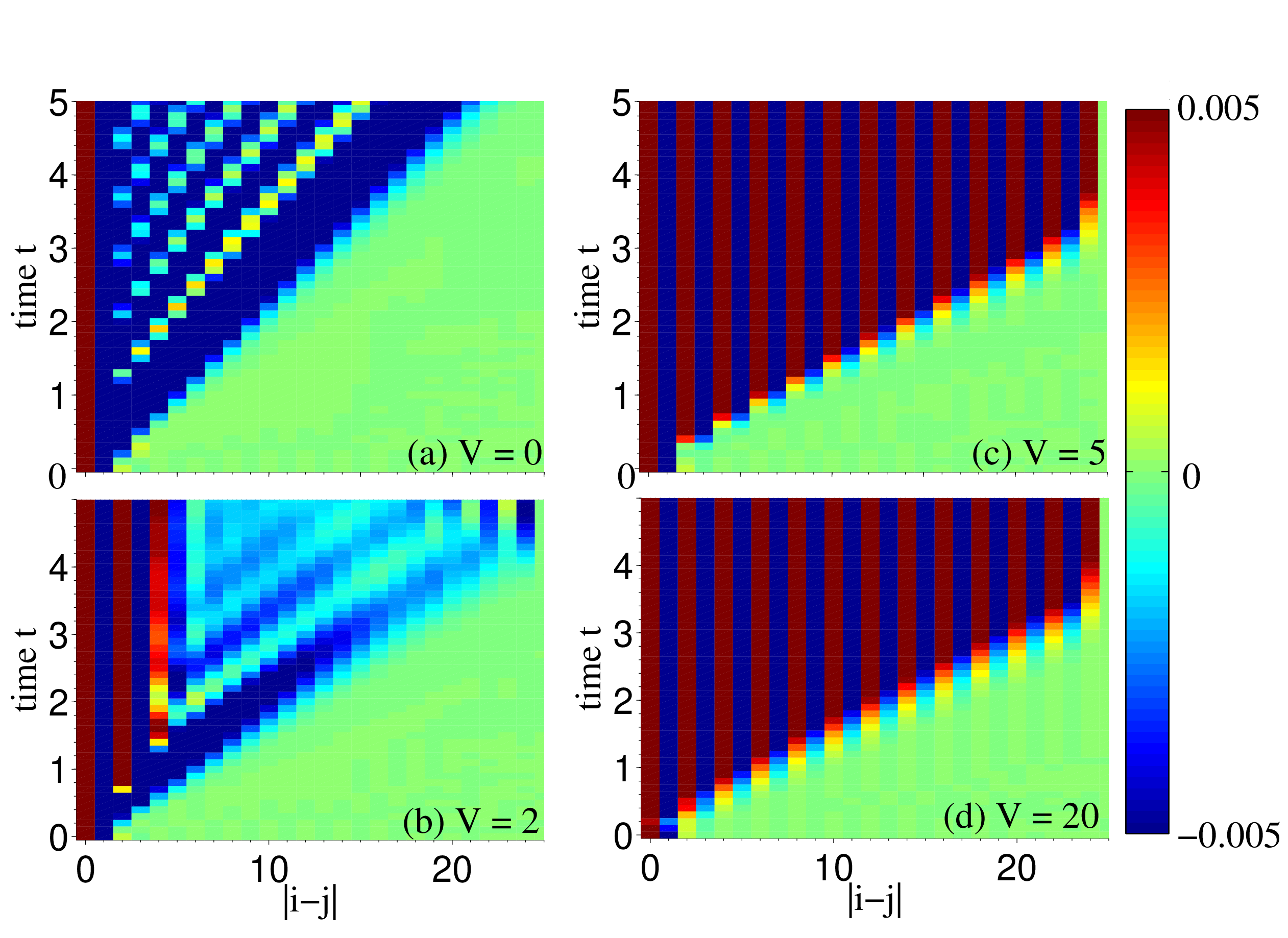}
\caption{(color online) Time evolution of the equal-time density correlation function
  $C_{i,j}(t)$ of  spinless fermions after a quench from the CDW
  ground state of $H(V_0$) with $V_0=10$, evolved by the
  Hamiltonian $H(V)$, with (a) $V=0$, (b) $V=2$, (c) $V=5$, (d) and $V=20$.}  
\label{fig:01}
\end{figure*}
We begin the discussion of our results by considering quenches that start from 
a CDW ground state, i.e., $H(V_0)$ with $V_0=10$. 
In Fig.~\ref{fig:01}, we provide a two-dimensional representation of the 
time evolution of the equal-time density correlation function
$C_{i,j}(t)$ for $V=0, \, 2, \, 5$, and $20$, corresponding to
Hamiltonians $H(V)$ with (noninteracting) Luttinger 
liquid ($V=0$), quantum critical ($V=2$), or CDW ($V=5, \, 20$) ground
states, respectively.
At time $t=0$, the density correlations decay exponentially with the separation 
$|i-j|$ due to the charge gap of the CDW ground state. 
After the quench, we observe that the time evolution of 
$C_{i,j}(t)$ exhibits a pronounced signature of a light cone in all cases. 
The local densities at two separated lattice sites remain essentially uncorrelated up to a time 
$t_d$, which increases linearly with their spatial separation $d=|i-j|$.  
The corresponding propagation velocity $u$ of the front can be obtained from the slope of 
the light cone using $t_d=d/u$. 
For systems describable by conformal field theory
and for various exactly solvable models, both in the continuum and on
discrete lattices, Calabrese and Cardy have shown that the velocity of this light-cone-like spread of the
correlations is twice the maximal group velocity $v$ of the
fastest excitations.
The relevant excitations in the present model of spinless fermions are the 
sound modes of the local density fluctuations. 
In particular, for the free case $V=0$, the slope of the first front
corresponds to $u(V=0)=2v_F=4t_{\text{h}}$, as expected, where $v_F$ denotes the Fermi velocity for $V = 0$.
In addition to the light cone, additional propagation fronts at later times
can be identified in Fig.\ \ref{fig:01}(a), which, however, possess a lower velocity.
This signals that slower quasiparticles stemming from regions without linear dispersion
also participate in spreading information.
Figure \ref{fig:01}(c) shows the evolution of the correlation function for a 
quench within the CDW phase, i.e., a case which should not be
describable by conformal field theory. 
Interestingly, we nevertheless find a pronounced light-cone behavior in the
correlation function.  
Although the conformal field theory underlying the treatment of Calabrese and Cardy is not valid
in this region, the physical picture that ballistically propagating quasiparticles 
are generated by the quench seems to hold.
However, in contrast to the case of the quench to the LL displayed in
Figs.\ \ref{fig:01}(a) and (b), 
we see that a strong alternating pattern forms in the density
correlation function and remains present and qualitatively unchanged
after the onset of the light cone.

\begin{figure}[ht]
\includegraphics[width=0.5\textwidth]{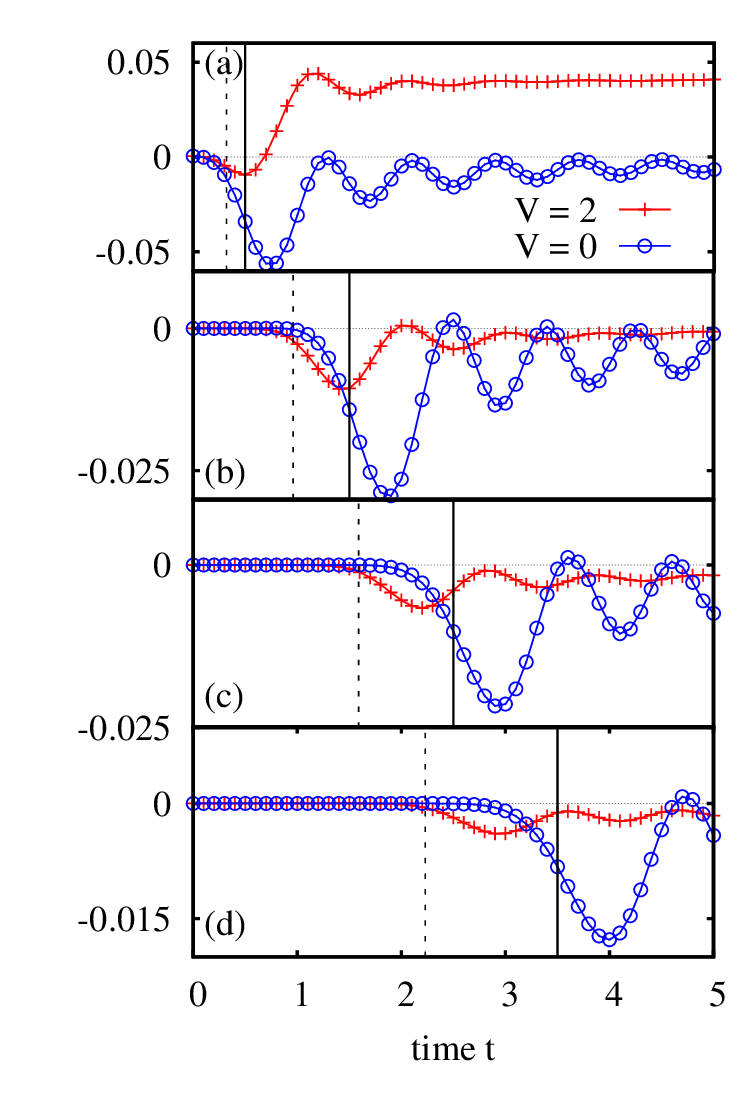}
\caption{(color online) Equal-time density correlation function $C_{i,j}(t)$ of
spinless fermions after a quench from the CDW ground state of $H(V_0$)
with $V_0=10$, evolved by  the Hamiltonian $H(V)$, with $V=0$ and
$V=2$, at fixed separations  (a) $|i-j|= 2$, (b) $6$, (c) $10$, and
(d) $14$ as functions of time $t$.  
The black vertical lines indicate the position of the horizon for $V=0$,
which moves with velocity $u = 4$, and
the dashed vertical lines a horizon 
moving with velocity $u=2 v(V=2) = 2 \pi$.}
\label{fig:02}
\end{figure}

A more detailed view of the temporal evolution of the correlation functions is shown 
in Fig.~\ref{fig:02}, in which we plot the values of $C_{i,j}(t)$ as a function 
of time for increasing distance $\mid i-j \mid$ for $V=0$ and $V=2$,
the two extremes of the Luttinger-liquid phase. 
After the arrival of the first signal, oscillatory behavior 
as a function of time can be observed at each distance.
However, as $V$ is increased, the observed
oscillations both decrease 
in magnitude and are damped out more rapidly. 
Comparing the results for the free case to 
the ones obtained for $V=2$ in Fig.~\ref{fig:02}, it can be seen that
the incoming front travels with a higher  velocity when $V$ is larger,
as can also be seen in Fig.\ \ref{fig:01}.

\begin{figure}[ht]
\includegraphics[width=0.5\textwidth]{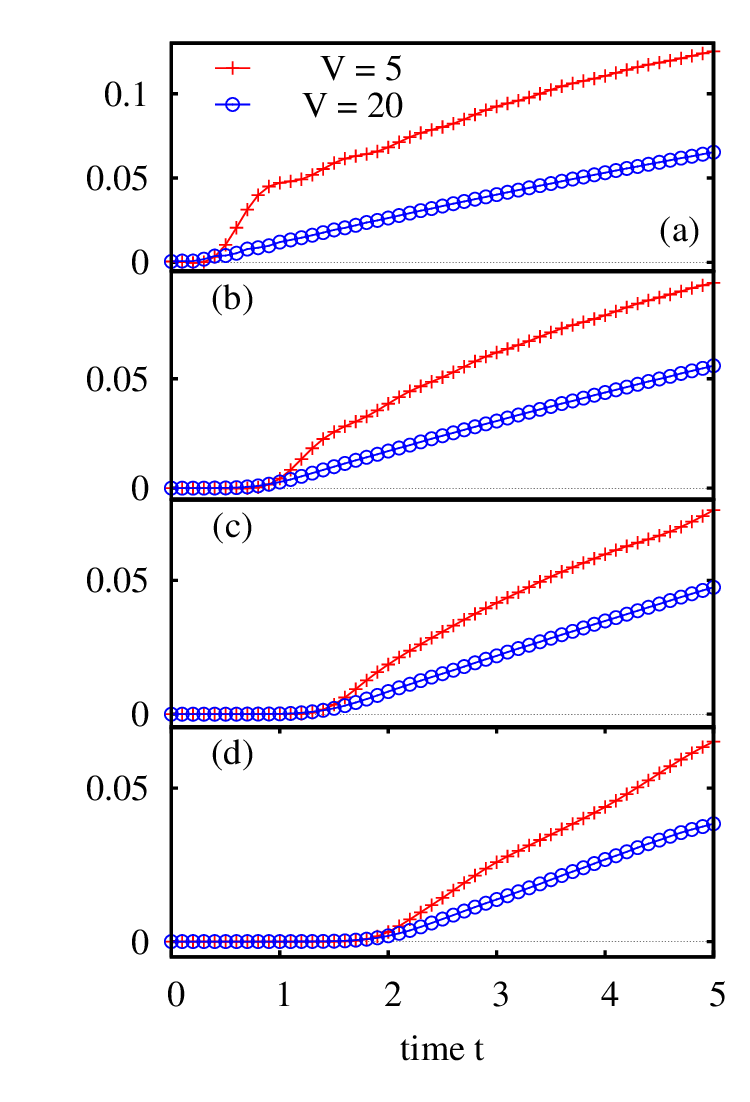}
\caption{(color online) Equal-time density correlation function $C_{i,j}(t)$ of
spinless fermions after a quench from the CDW ground state of $H(V_0$)
with $V_0=10$, evolved by  the Hamiltonian $H(V)$, with $V=5$ and
$V=20$, at fixed separations  $|i-j|= 2$ (a), $6$ (b), $10$ (c), and
$14$ (d) as functions of time $t$ after the quench.}
\label{fig:03}
\end{figure}

In contrast to the oscillatory behavior in the Luttinger-liquid phase,
a steady increase of the correlations is observed when the quench
occurs within the CDW phase, as can be seen in Fig.\ \ref{fig:03}.
The alternating pattern imprinted at the onset of the light cone is preserved.
Presumably, the correlation functions saturate at some time that is
significantly longer than the maximum time reached here.
While results for both $V < V_0$ and $V > V_0$ show the same
qualitative behavior, a difference is observed in the rate and the
form of the increase of the correlation functions.
The increase of the data is stronger for $V=5$ in
Fig.\ \ref{fig:03} but is sublinear, while the increase is linear to a
very good approximation for $V=20$. 

\begin{figure}[th]
\includegraphics[width=0.5\textwidth]{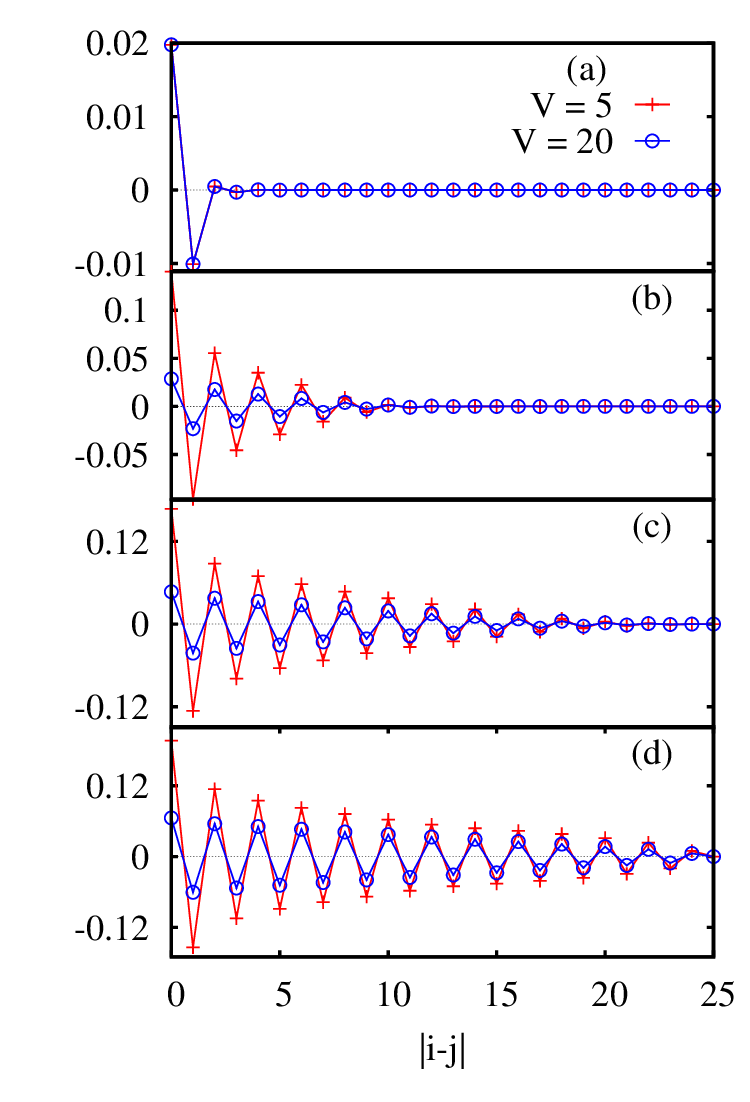}
\caption{(color online) Equal-time density correlation function $C_{i,j}(t)$ after a
  quench from the CDW ground state of $H(V_0$) with $V_0=10$, 
  evolved by  the Hamiltonian $H(V)$, with $V=5$ and $V=20$, at fixed
  times (a) $t=0$, (b) $t=1.4$, (c) $t=2.8$, and (d)
  $t=4.2$ (d) as functions of the separation $|i-j|$.} 
\label{fig:05}
\end{figure}

\begin{figure}[th]
\includegraphics[width=0.475\textwidth]{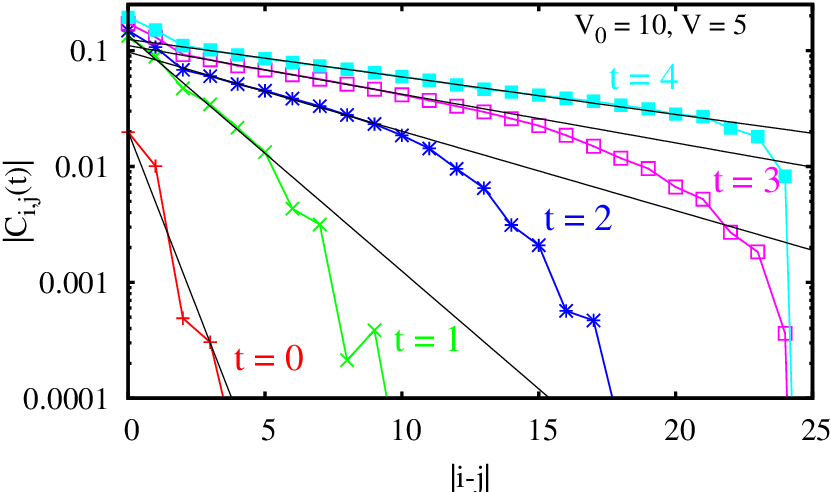}
\caption{(color online) Equal-time density correlation function as a function of
  distance $|i-j|$ at different times for the quench from
  $V_0 = 10$ to $V = 5$. The black lines are fits to exponential decays.} 
\label{fig:growingcorrelations}
\end{figure}

This increase of correlations is remarkable.
The length over which the correlation function falls off is much
larger than the initial correlation length  after the quench, as can
be seen in Fig.\ \ref{fig:05} by comparing the initial $C_{i,j}(t)$
[Fig.\ \ref{fig:05} (a)] with the results for later
times. 
This is analyzed in more detail in Fig. \ref{fig:growingcorrelations}
for the quench from $V_0 = 10$ to $V = 5$. 
As can be seen on the semilogarithmic scale, $|C_{i,j}(t)|$ 
decays exponentially for intermediate distances at all times 
$t \leq 5$.
The correlation length and the distance over which the decay is
exponential both increase substantially with time.
This can be contrasted to the results of Ref.\ \onlinecite{cazalilla06}
for the Luttinger 
model where, starting from a gapless phase, an algebraic decay of the
correlations is found.

\begin{figure}[th]
\includegraphics[width=0.475\textwidth]{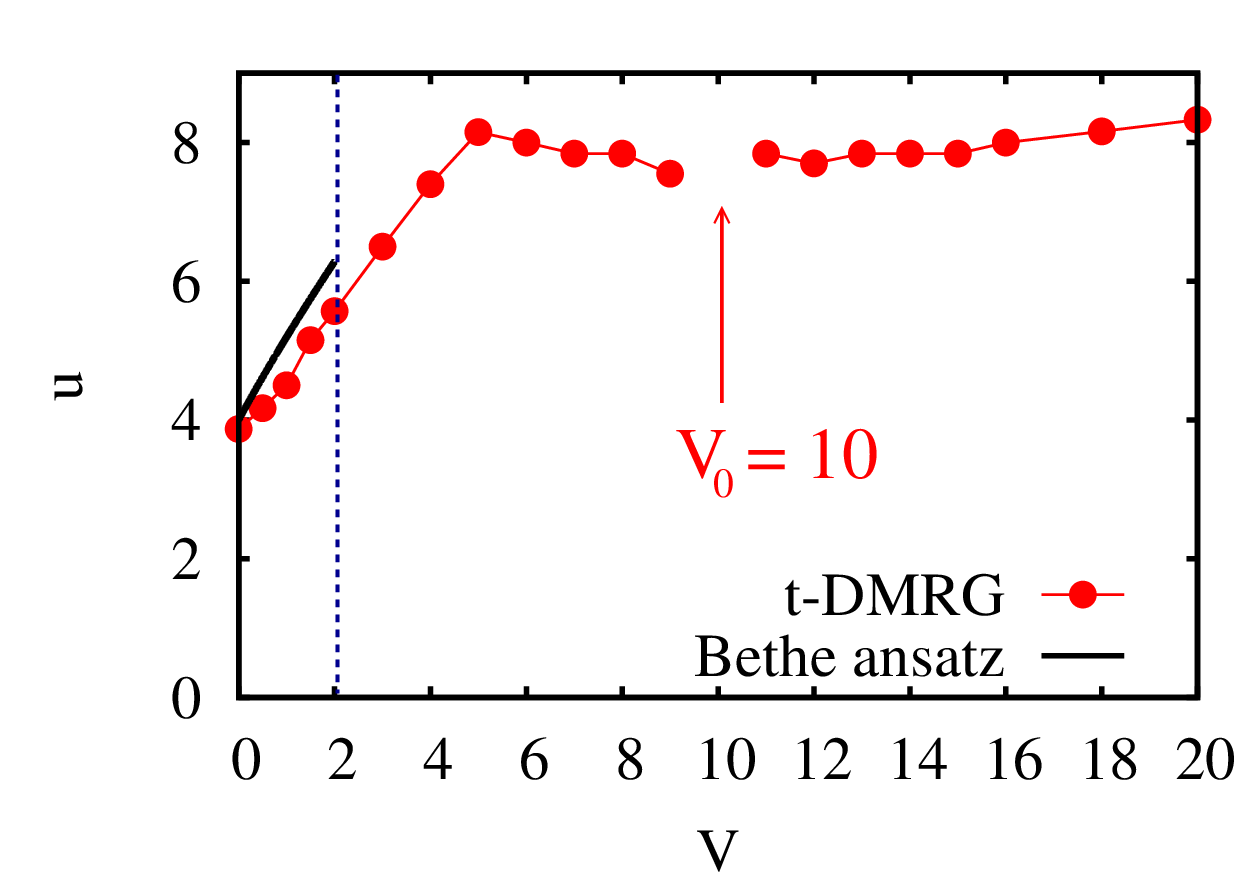}
\caption{(color online) Velocity of the the horizon 
in the time evolution of the 
equal-time density correlation function $C_{i,j}(t)$ of
spinless fermions after a quench from the CDW ground state of $H(V_0$)
with $V_0=10$,  evolved by the Hamiltonian $H(V)$ as a function of
$V$.  
The estimated error is of the order of the point size.
The dashed vertical line indicates the quantum phase transition form
the Luttinger liquid  to the CDW regime. 
}
\label{fig:06}
\end{figure}

Note that the characteristic velocity of the light cone increases as
$V$ is changed from $V=0$ to $V=2$, as can be seen in
Figs.\ \ref{fig:01} (a) and (b) and in Fig.\ \ref{fig:02}, 
but is approximately the same for $V=5$ and $V=20$,
Figs.\ \ref{fig:01} (c) and (d) and 
Fig.\ \ref{fig:03}. 
The increase of velocity in the Luttinger liquid phase can be 
understood on the basis of  the Bethe ansatz solution of the
equivalent spin-1/2 
XXZ model,\cite{gap_Bethe} where the relevant  maximum mode velocity 
in the Luttinger liquid regime $V<2$ can be obtained as \cite{review_Schulz}
\begin{equation}
v(V) = v_F \frac{\pi \sqrt{1 - (V/2)^2}}{2 \, {\arccos}(V/2)},
\label{eq:velocities}
\end{equation}
where in our case $v_F = 2 t_{\text{h}}$. 
It can be seen that the front propagating with velocity $u=2v(V)$
indeed coincides with the shoulder of the first peak in the signal of
$C_{i,j}(t)$ in Fig.\ \ref{fig:02}.
We display the light-cone velocities for different values of $V$
in Fig.\ \ref{fig:06}, together with the result from the Bethe ansatz. 
We see that inside the Luttinger liquid regime, $V<2$, the horizon travels at a 
velocity $u(V)$ that approximately follows the Bethe-ansatz
prediction $2 v(V)$, but is consistently somewhat smaller, except at
$V=0$.
We thus confirm the general picture obtained by Calabrese and Cardy of ballistic
propagation of information through the system, but find that the
propagation does not take place with the maximal possible velocity.
On general grounds, conformal field theory becomes exact in the
asymptotic limit of a continuum theory in which the quasiparticles
have a linear dispersion relation.
However, after the quench, quasiparticles are
excited over a broad energy range and hence generally have velocities
lower than the maximum set by Eq.~\ref{eq:velocities}.
We attribute the reduced apparent light-cone propagation velocity
found in Fig.~\ref{fig:06} to the contribution of these lower-velocity
quasiparticles.

For values of the interaction beyond the quantum critical point, the 
velocity of the horizon can still be determined, as is evident in
Figs.\ \ref{fig:01} (c) and (d). 
This velocity continues to increase up to $V \approx 5$. 
Beyond this value, $u(V)$ remains essentially 
constant, as could already be seen in Fig.\ \ref{fig:03}.

We conclude this section by comparing the commensurability of the correlations 
after quenches that remain in the CDW phase to that of ones
quenching to the LL phase.
The presence of a CDW ground state for $V_0=10$ leads to strong commensurate
spatial oscillations in $C_{i,j}(t)$ which increase in time, as demonstrated for $V=5$ and
$V=20$ in Fig.~\ref{fig:05}.
For quenches beyond the insulating regime, however, we also observe spatial 
oscillations in $C_{i,j}(t)$, but with larger spatial periods incommensurate to 
the underlying lattice (see Fig.~\ref{fig:01}). 
In the propagating quasiparticle picture of Calabrese and Cardy, 
such spatial oscillations are expected to arise in the 
time evolution of correlation functions as a generic property of a system 
with a nonlinear dispersion relation.
It is rather remarkable that for interaction quenches that stay within the CDW regime, 
the time evolution leads to the buildup of 
extended commensurate correlations
even though the ground states in this insulating regime possess a very fast
exponential decay of the very same correlation function. 

\subsection{Luttinger-liquid initial state}

\begin{figure}[th]
\includegraphics[width=0.5\textwidth]{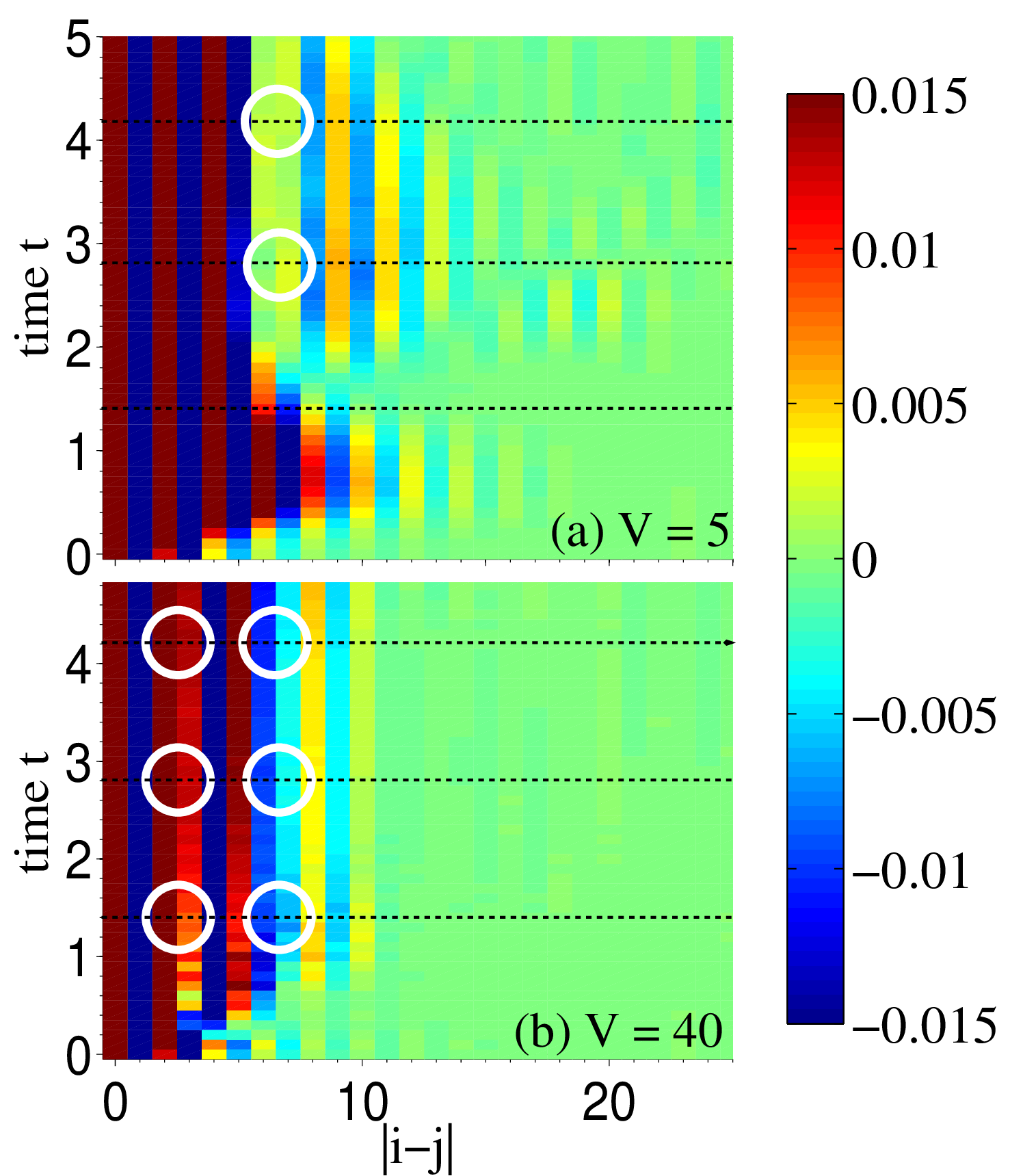}
\caption{(color online) Time evolution of the equal-time density correlation function
  $C_{i,j}(t)$ after a quench from the Luttinger-liquid ground state
  of $H(V_0$) with $V_0=0.5$, evolved by the Hamiltonian $H(V)$,
  with (a) $V=5$ and (b) $V=40$.
  The dashed horizontal lines indicate the time steps shown in Figs.~\ref{fig:09}(b)-(d); the white circles highlight the phase slips visible there.}
\label{fig:07}
\end{figure}

We now consider quenches that start from the Luttinger-liquid ground
state of $H(V_0)$, taking $V_0=0.5$.  
We focus particularly on quenches that go beyond the quantum 
critical point to the symmetry-broken CDW regime. 
In Fig.\ \ref{fig:07}, we 
show the time evolution of the equal-time density correlation function  
$C_{i,j}(t)$ in a two-dimensional representation for $V=5$ and $V=40$. 
At $t=0$, the density correlations $C_{i,j}(0)$ decay algebraically
due to the critical nature of the $V_0=0.5$ ground state. 
For quenches with $V \lesssim 5$, a light cone can be identified, as 
seen in Fig.~\ref{fig:07}(a) for $V = 5$, 
where it is visible for times $t \lesssim 0.5$. 
Bearing in mind 
that the predictions of Calabrese and Cardy are based on
an initial state with a finite correlation length,
this finding is somewhat surprising. 
The slow algebraic decay of the correlations in the initial state
causes the central site and 
the site at the boundaries of the system to already be correlated at $t = 0$.
Therefore, the propagation of entangled quasiparticles created by the
quench at the open boundary of the system leads to an additional
signal which appears at decreasing values of $\mid i-j \mid$ when $t$ is increased.
This can be seen in Fig.~\ref{fig:07}(a) for sufficiently small values of $|i-j| \lesssim 10$
for times $1 \lesssim t \lesssim 2$, where an additional light cone moves towards the left
at a velocity approximately half of the velocity of the main light cone.
This observation is in agreement with the picture of Calabrese and Cardy applied to a
semi-infinite chain: the left-moving signal in $C_{i,j}(t)$ is caused
by the ballistic propagation of a pair of quasiparticles created
during the quench, one of which is reflected at  
the open chain boundary.

\begin{figure}[t]
\includegraphics[width=0.5\textwidth]{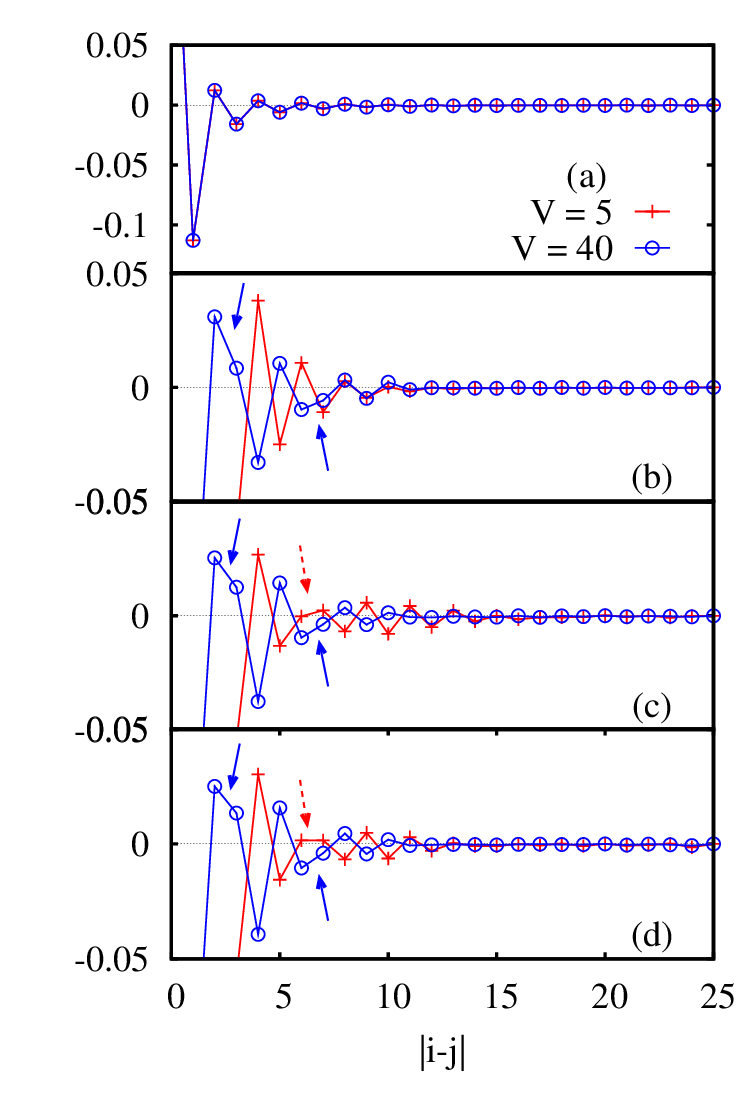}
\caption{(color online) Equal-time density correlation function $C_{i,j}(t)$ of
spinless fermions after a quench from the Luttinger liquid  ground
state of $H(V_0$) with $V_0=0.5$, evolved by  the Hamiltonian
$H(V)$, with $V=5$ and $V=40$, at fixed  times $t=0$ (a), $t=1.4$ (b),
$t=2.8$ (c), and $t=4.2$ (d) as functions of the separation $|i-j|$.
The blue (solid) arrows highlight the position of the phase slips obtained for $V = 40$, the red (dashed) ones point to the phase slips 
for the case $V = 5$.} 
\label{fig:09}
\end{figure}

We now turn to quenches deep into the CDW regime.
In Fig.~\ref{fig:07}(b), our results for the quench from $V_0 = 0.5$
to $V = 40$  are displayed.
It is not possible to identify the signature of a light cone
in the data.
Here $C_{i,j}(t)$ does not change significantly for times 
$t \gtrsim 1/t_{\text{h}}= 1$, the typical time scale of a single fermion
hopping process.
In contrast to the previous cases, 
phase slips, marked by a reversal of the phase of the alternation of
the correlations, are present.
These phase slips are pronounced in $C_{i,j}(t)$ for $V = 40$, but
can also be seen for $V = 5$ at distances $\mid i-j \mid =6-7$ for times $t \gtrsim 2$.
This can be seen in more detail in Fig.~\ref{fig:09},
where the phase slips for $V = 40$ are clearly visible in Figs.~\ref{fig:09}(b)-(d)
(corresponding to times $t > 1.4$), while for $V = 5$ they only appear
in Figs.~\ref{fig:09}(c) and (d).

\begin{figure}[b]
\includegraphics[width=0.5\textwidth]{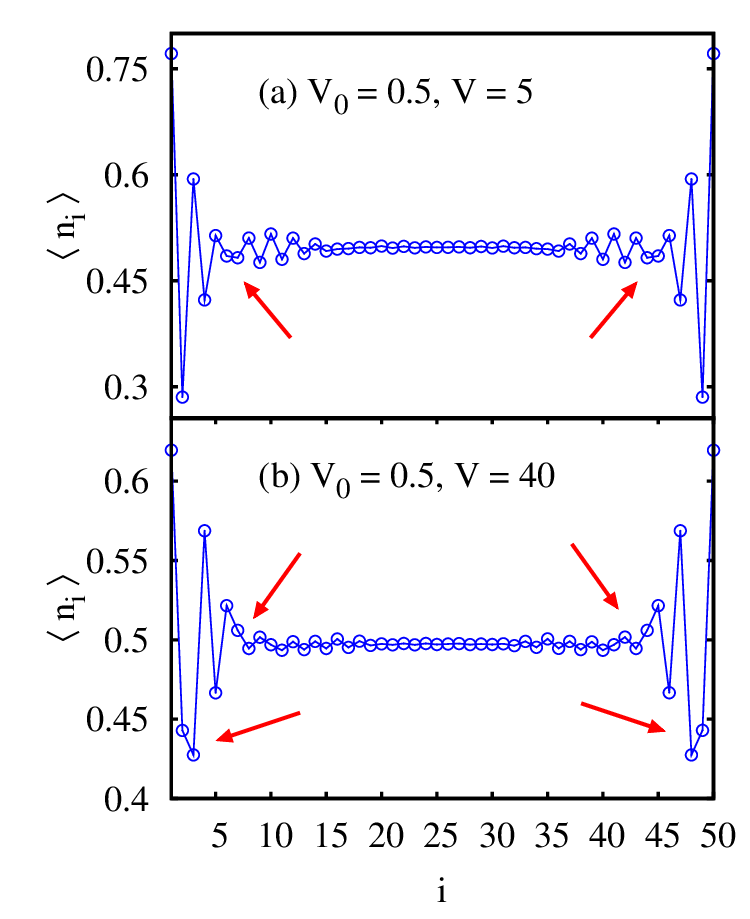}
\caption{(color online) Local density $\langle n_i \rangle(t)$ for a system of length
  $L=50$ with $N=25$ particles
  at time $t = 5$ after a quench from the Luttinger liquid  ground
  state of $H(V_0)$ with $V_0 = 0.5$, evolved by the Hamiltonian
  $H(V)$, with (a) $V = 4$ and (b) $V = 40$. 
  Arrows indicate the location of domain walls (kinks) in the local CDW pattern.} 
\label{fig:10}
\end{figure}

The presence of such phase slips in the correlation function is not
observed in the quenches starting from a CDW state which we discussed above,
and suggests the presence of different CDW domains in the local
density distribution which are separated by domain walls. 
Therefore, we now analyze the time evolution of the local density 
$\langle n_i \rangle(t)$ directly.
Results for the local density distribution at $t=5$ are shown for both
$V=5$ and $V=40$ in Fig.~\ref{fig:10}. 
We find that domain walls (kinks) which separate different regions of
CDW modulations and are stable in time are formed in the vicinity of
the boundaries. 
Comparing the two values of $V$, we also find that the number of such
domain walls is larger for larger values of $V$ (at $t=5$ we identify  
2 such kinks for $V=5$, and 4 for $V=40$).
Note that the density $\langle n_i \rangle(t)\approx 0.5$ in the
middle of the chain shows almost no oscillation for this exactly
half-filled lattice ($L=50$, $N=25$).
The open ends of the chain act as effective impurities and induce
Friedel-like oscillations in the local density, which fall off
similarly to the correlation function.
Therefore, the 
stability of the domains in 
$\langle n_i \rangle(t)\approx 0.5$ reflects the behavior of the phase
slips in the density correlation function.
The 
presence of the phase slips and domains 
is reminiscent
of the scenario proposed by  Kibble \cite{kibble76} and Zurek
\cite{zurek85,zurek96} concerning the generation of topological defects in
quenches that cross a quantum phase transition point\cite{zurek:105701} to a regime with
spontaneous symmetry breaking. 
In the corresponding Kibble-Zurek mechanism, the duration of the quench
$\tau$ is usually assumed to be finite. 
The characteristic critical scaling behavior near the crossed critical point
leads to a power-law scaling of the defect density with $1/\tau$.
In the current situation, the quench is sudden, corresponding 
to $\tau=0$.
The generation of defects in our system is therefore well 
beyond the Kibble-Zurek scaling regime.
Pellegrini {\em et al.} have studied the scaling of the energy excess
with duration in an XXZ spin chain undergoing a 
quench through the critical point.\cite{pellegrini:140404}
It would be interesting to study directly the dependence of the
defect density on the duration of the quench for this model in order to link our
results for the instantaneous quench with the scaling prediction of
the Kibble-Zurek mechanism.

\subsection{Entanglement entropy of subsystems}

\begin{figure}[th]
\includegraphics[width=0.5\textwidth]{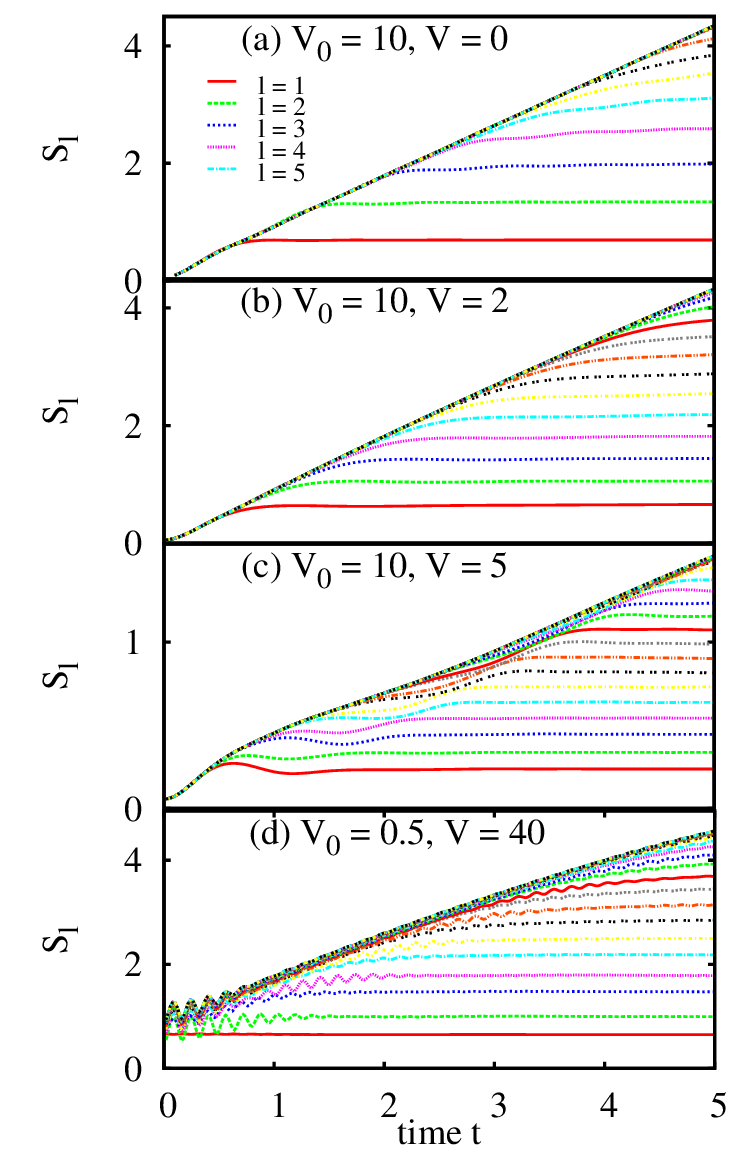}
\caption{(color online) Time evolution of the von Neumann entropy $S_l(t)$ for
  subsystems of different sizes $l = 1, \, 2, \, \ldots, 25$, after
  quenches with (a) $V_0 = 10 \rightarrow V = 0$, 
  (b) $V_0 = 10  \rightarrow V = 2$, (c) $V_0 = 10 \rightarrow V = 5$,
  and (d) $V_0 = 0.5 \rightarrow V = 40$. 
  The lowest line corresponds to $l=1$, the next to $l=2$ and so on;
  for clarity, the curves for $l \leq 5$ are labeled.}
\label{fig:11}
\end{figure}

In the previous sections we have characterized the propagation of
information after a quench by the behavior of the equal-time density
correlation function. 
We find two different scenarios: 
{\em (i)} information spreads ballistically through the system and
leads to a light cone, and 
{\em (ii)} domain walls form after a quench deep into the
symmetry broken regime,  
reminiscent of the Kibble-Zurek scenario.

The specific quantum nature of the problem under consideration can be
further analyzed by directly investigating the entanglement buildup
and spread in the system after the quench.
A direct measure of the entanglement inside the system is given by the von Neumann entropy $S_l$ of subsystems
of size $l$ as defined in Eq.~(\ref{eq:blockentropy}).
In Fig.\ \ref{fig:11}, we show t-DMRG results for the time
evolution $S_l(t)$ for some of the quenches considered previously. 
In all cases, we find that the block entropy for a particular
subsystem size eventually saturates with
time, to a final value that increases linearly with subsystem size.
The initial phase of the time evolution of $S_l(t)$, however, shows 
characteristic differences for the two cases considered above: For those 
cases in which a distinct light cone has been observed in the density correlation 
function, the von Neumann entropy initially grows linearly, as seen
in Fig.~\ref{fig:11} (a) and (b) for quenches  
from $V_0=10$ to $V=0$ and $V=2$, respectively. 
There is therefore a characteristic time $t_l$ at which the entropy saturates.
One finds, to a good approximation, that $t_l$ increases linearly with the 
subsystem length $l$, giving rise to a characteristic entropy propagation 
velocity that is roughly equal to the light-cone velocity $u$ extracted from 
the density correlation function. 
The evolution of $S_l(t)$ in Fig.~\ref{fig:11}(c), corresponding to a quench 
from $V_0=10$ to $V=5$, shows more complex initial behavior, with the 
formation of intermediate plateaux in $S_l(t)$, e.g., for $l=5$, and the 
presence of local minima (dips) in $S_l(t)$, e.g., for $l=3$. 
Turning now to Fig.~\ref{fig:11}(d), a quench from a LL initial state with 
$V_0=0.5$ to a point deep in the symmetry-broken regime, $V=40$, we find
a broad sublinear increase rather than a linear growth in
$S_l(t)$, which, in addition, shows pronounced oscillations that damp
out towards saturation.
Similar oscillations were observed in Ref.\ \onlinecite{addendum1} and
were found to be due to the existence of additional local maxima of the group velocity of the quasiparticles.
Thus, the behavior of $S_l(t)$ further corroborates
the qualitative difference between 
the cases with ballistic transport, showing a clear light-cone effect
in the density correlation function $C_{i,j}(t)$,
and cases with no distinct light cone.
An interesting topic for future studies would be to analyze how the 
presence of kinks and phase slips in the local density and the density
correlation function relates to this peculiar sublinear rise of the
von Neumann entropy  in more detail, 
and to understand the origin of the pronounced initial oscillations.  

\section{Summary and Discussion}
\label{sec:discussion}

In this work, we have studied the time evolution of correlation
functions in a one-dimensional, half-filled system of spinless fermions with
nearest-neighbor repulsive interactions.
We perturb the system by changing the strength of the repulsion
suddenly, i.e., we quench the interaction parameter and study the
evolution of the density correlations at short to intermediate times.
The investigations were carried out using the adaptive time-dependent
density matrix renormalization group (t-DMRG).
Since the half-filled system has a gapless, metallic
(Luttinger-liquid) ground-state 
phase for small interaction strength and a gapped charge-density-wave
insulating phase for large interaction strength, separated by a quantum
phase transition point, 
it is possible to carry out quenches
both within the two qualitatively very different phases and from one phase
to another.
When we do this, we find that quenches starting from the CDW
insulating phase, either staying within the phase or quenching to the
metallic phase, are both characterized by a light cone in the density
correlation function, i.e., a horizon propagating with constant
velocity, beyond which changes  at a particular
separation become visible.
For the metallic case, the behavior agrees well with predictions based
on conformal field theory made by Calabrese and
Cardy, \cite{calabrese06,calabrese07} although there appear to be
small corrections to a horizon velocity obtained from the exactly
known velocity of charge excitations.

For quenches within the insulating phase, the presence of such a clear
horizon for all values of the final interaction strength is a
surprise.
One might expect a horizon to occur when the energy scale of the gap
is smaller than the excess energy (relative to the ground state)
associated with the quench, but this does not seem to be the case.
While there is presumably some sort of ballistically propagating
quasiparticle associated with the horizon, its nature is not yet
clear.
In addition, the quench within the insulator is characterized by 
a density correlation function which remains exponentially decaying,
but with a correlation length which grows in time, at least to the
time scales that could be reached numerically.
The behavior of the correlation length with time at longer times
remains to be clarified.

For quenches from a metallic initial state to an insulating parameter
value, the presence or absence of the remnants of a light cone depend
on the strength of the final interaction.
For relatively weak final interaction, a light cone is present to a
time scale characterized by the propagation time to the end of the
chain.
Counter-propagating light cones with one-half the velocity of the
initial light cone are also present.
The counter-propagating features can be understood in terms of quasiparticles
propagating in one direction only from the open ends of the chain.
Examination of the correlation function reveals phase slips in the
correlation functions and the formation of domains in the local
density, which are indications of frozen-in domains generated by the
quench.
Such domains are predicted by the Kibble-Zurek
mechanism,\cite{kibble76,zurek85,zurek96} which seems 
to be applicable here even though our sudden quench falls outside its
scaling regime.

The behavior of the quantum information (von Neumann) entropy of
subsystems supports the picture obtained from the behavior of the
correlation functions.
For cases where a light cone with a constant velocity is present in
the correlation functions, there is a linear growth of the entropy
with time up to a saturation value associated with the system size.
A velocity associated with the saturation time is consistent with the
light-cone velocity.
For the cases with no light cone, the growth of the entropy with time
is sublinear, with no well-defined saturation time.
In addition, oscillatory behavior, which could be interesting to
understand in more detail, occurs.

\section*{Acknowledgements}
We thank A.\ L\"auchli, C.\ Kollath, S.\ Montangero, and
F.\ Gebhard for useful discussions.
We acknowledge HLRS Stuttgart and NIC J\"ulich for allocation of CPU
Time. 
The SFB/TR 21 is acknowledged for financial support.

\end{document}